\documentstyle[11pt,amssymb,epsfig,amsmath]{article}

\begin{document}
\newcommand{\be}{\begin{equation}}
\newcommand{\ee}{\end{equation}}
\newcommand{\ba}{\begin{eqnarray}}
\newcommand{\ea}{\end{eqnarray}}
\newcommand{\no}{\nonumber \\}
\newcommand{\gsim}{\mathrel{\hbox{\rlap{\lower.55ex \hbox {$\sim$}}
                   \kern-.3em \raise.4ex \hbox{$>$}}}}
\newcommand{\lsim}{\mathrel{\hbox{\rlap{\lower.55ex \hbox {$\sim$}}
                   \kern-.3em \raise.4ex \hbox{$<$}}}}

\def\be{\begin{eqnarray}}
\def\ee{\end{eqnarray}}
\def\bea{\be}
\def\eea{\ee}
\newcommand{\e}{{\mbox{e}}}
\def\del{{\partial}}
\def\vr{{\vec r}}
\def\vk{{\vec k}}
\def\vq{{\vec q}}
\def\vP{{\vec P}}
\def\vt{{\vec \tau}}
\def\vs{{\vec \sigma}}
\def\vJ{{\vec J}}
\def\vB{{\vec B}}
\def\hatr{{\hat r}}
\def\hatk{{\hat k}}
\def\roughly#1{\mathrel{\raise.3ex\hbox{$#1$\kern-.75em%
\lower1ex\hbox{$\sim$}}}}
\def\lsim{\roughly<}
\def\gsim{\roughly>}
\def\fm{{\mbox{fm}}}
\def\vx{{\vec x}}
\def\vy{{\vec y}}
\def\({\left(}
\def\){\right)}
\def\[{\left[}
\def\]{\right]}
\def\EM{{\rm EM}}
\def\barp{{\bar p}}
\def\zz{{z \bar z}}
\def\mus{{\cal M}_s}
\def\abs#1{{\left| #1 \right|}}
\def\ve{{\vec \epsilon}}
\def\nlo#1{{\mbox{N$^{#1}$LO}}}
\def\MS{{\mbox{M1V}}}
\def\mut{{\mbox{M1S}}}
\def\Qt{{\mbox{E2S}}}
\def\rM{{\cal R}_{\rm M1}}\def\rE{{\cal R}_{\rm E2}}
\def\la{{\Big<}}
\def\ra{{\Big>}}
\def\lsim{\mathrel{\rlap{\lower3pt\hbox{\hskip1pt$\sim$}}
     \raise1pt\hbox{$<$}}} 
\def\gsim{\mathrel{\rlap{\lower3pt\hbox{\hskip1pt$\sim$}}
     \raise1pt\hbox{$>$}}} 
\def\N{${\cal N}\,\,$}

\def\af{\alpha}
\def\bt{\beta}
\def\baf{{\bar\alpha}}
\def\bbt{{\bar\beta}}
\def\eL{\epsilon_L}
\def\eR{\epsilon_R}
\def\xiL{\xi^L}
\def\xiR{\xi^R}
\def\xL{\chi^L}
\def\xR{\chi^R}
\def\Aq{A_{em}}
\def\Dq{{D^Q}}
\def\ka{\kappa}
\def\lam{\lambda}
\def\Lam{\Lambda}
\def\Gm{\Gamma}
\def\dlt{\delta}
\def\eps{\epsilon}
\def\sig{\sigma}
\def\omg{\omega}
\def\Omg{\Omega}
\def\vp{{\vec \xi}}
\def\vu{{\vec u}}
\def\ve{{\vec E}}
\def\vb{{\vec B}}
\def\crs{\times}
\def\lv{\lvert}
\def\rv{\rvert}

\def\J#1#2#3#4{ {#1} {\bf #2} (#4) {#3}. }
\def\PRL{Phys. Rev. Lett.}
\def\PL{Phys. Lett.}
\def\PLB{Phys. Lett. B}
\def\NP{Nucl. Phys.}
\def\NPA{Nucl. Phys. A}
\def\NPB{Nucl. Phys. B}
\def\PR{Phys. Rev.}
\def\PRC{Phys. Rev. C}

\renewcommand{\thefootnote}{\arabic{footnote}}
\setcounter{footnote}{0}

\begin{flushright}
MPP-2011-150
\end{flushright}

\vskip 0.4cm \hfill { }
 \hfill {\today} \vskip 1cm

\begin{center}
{\LARGE\bf An anomalous hydrodynamics for chiral superfluid
   }
\date{\today}

\vskip 1cm {\large Shu Lin
\footnote{E-mail: slin@mpp.mpg.de}
\\Max-Planck-Institut f\"{u}r Physik (Werner-Heisenberg-Institut)
\\ F\"{o}hringer Ring 6, 80805 M\"{u}nchen, Germany}


\end{center}

\vskip 0.5cm

\begin{center}

\end{center}

\vskip 0.5cm

\begin{abstract}

Starting from low energy effective chiral Lagrangian with gauged Wess-Zumino Witten term, we have derived a hydrodynamic theory for chiral superfluid. It is a non-abelian hydrodynamics at zero temperature with only superfluid components. With an external electromagnetic field and baryonic and axial baryonic chemical potentials turned on, we are able to identify analogs of various anomaly induced term in normal hydrodynamics, including chiral vortical effect, chiral magnetic effect and chiral electric effect. As an example, we solved the hydrodynamic equations for the ground state and observed the chiral magnetic effect and chiral separation in the confined phase.

\end{abstract}

\newpage

\renewcommand{\thefootnote}{\#\arabic{footnote}}
\setcounter{footnote}{0}

\section{Introduction}

Successful applications of the fluid/gravity duality \cite{flugra} to the R-charge black hole background \cite{haack,indian} have found a vorticity induced contribution to the constitutive equation of the R-current. The vorticity induced term originating from a Chern-Simon term in the gravity description encodes chiral anomaly in the dual field theory. It signals that quantum effect like chiral anomaly can appear as parity odd terms in macroscopic hydrodynamics. It was later shown in a seminal paper by Son and Sur\'owka \cite{surowka} that parity odd terms are actually required by second law of thermodynamics for hydrodynamics with chiral anomaly: By demanding the existence of a entropy current with non-negative divergence, they were able to fix all the transport coefficients of the parity odd terms, which include chiral magnetic effect (CME)\cite{kharzeev} and chiral vortical effect (CVE)\cite{KS}, in the constitutive equations of currents. Due to the non-renormalization of the anomaly coefficients, the anomaly induced effect are robust. There have been extensive model studies at both strong coupling using gauge/gravity duality \cite{rebhan,zayakin,kirsch,obannon,yee,neiman} and at weak coupling using field theory \cite{pu,hou}. Recently, the entropy current principle was also extended to higher dimension \cite{logna} and higher order hydrodynamics \cite{KY}.

It was pointed out in \cite{oz} that the transport coefficients are fixed by the entropy current principle only up to unknown functions of temperature. It has been established later that the unknown functions are related to the gravitational anomaly \cite{landsteiner}. The situation becomes more complicated when the system contains spontaneous symmetry breaking. The Goldstone bosons resulting from the symmetry breaking is gapless, thus have to be introduced to the hydrodynamics as additional degrees of freedom, leading to superfluid hydrodynamics. the case of $U(1)$ superfluid hydrodynamics has been studied in \cite{lin,Min2,oz2}, based on early works on non-anomalous $U(1)$ superfluid hydrodynamics \cite{Min,Yarom}. The dynamics of the Goldstone boson enters the constitutive equations as superfluid velocity. The presence of the superfluid velocity lowers the predictive power of the entropy current principle, leaving several transport coefficients as arbitrary functions of thermodynamical quantities.

There have been also effective field theory approaches \cite{lublin,nicolis} to derive the constitutive equations of hydrodynamics. Since the anomaly arises as gauge variation of the Wess-Zumino-Witten (WZW) term, the effect of anomaly can be incorporated by introducing a WZW term to the effective field theory action. One example is the chiral perturbation theory as the low energy effective field theory of QCD vacuum. The WZW term corresponding to the chiral anomaly in this case is explicitly known. The combined action can be used to study the anomalous effect on the chiral superfluid hydrodynamics. This is essentially the approach taken in \cite{lublin}. A different approach has been taken in \cite{nicolis}, where the authors wrote down the effective field theory for fluid, and constructed the most general WZW term in dimension two. The resulting action allowed them to determine the anomalous transport coefficient explicitly. In this paper, we will follow the lines of \cite{lublin}. We would like to emphasize here that since the chiral Lagrangian contains only Goldstone bosons as the fundamental field, what we will obtain is hydrodynamics with only superfluid components with non-abelian group structure. Our work distinguishes from \cite{lublin} in many fine details as will be clear in what follows. These lead to quantitative differences in the resulting constitutive equations of the currents.

This paper is organized as follows: In Section 2, we will focus on the abelian currents. We will derive the consistent current from the WZW term, and pass to covariant current by adding proper local terms in the external field. Section 3 will be a parallel study on the non-abelian currents. In Section 4, we will define properly the hydrodynamic variables and formulate a chiral superfluid hydrodynamics. Many interesting structures emerge from the WZW term, including the superfluid analogs of CME, CVE and also chiral electric effect (CEE)\cite{oz2}. Section 5 includes an application of the hydrodynamics to the ground state of QCD, where we find the chiral magnetic effect and chiral separation effect. We will conclude in Section 6.

\section{Abelian currents from gauged WZW term}

Our starting point is chiral Lagrangian plus gauged WZW term\cite{wz,witten}. Since it is a low energy effective action, we will work in the mean field approximation, where all fields take their classical values. We will follow mostly the notations of \cite{KRS} in this paper. The full action is given by:
\begin{align}\label{f_S}
&S=\int d^4x{\cal L}_0+\Gm_{WZ}(U,V,A) \\
&{\cal L}_0=-\frac{f^2}{8}Tr(D_\mu U D^\mu U^+),
\end{align}
where $U=e^{\frac{2i}{f}\phi}$ is the unitary matrix parameterized by the 
pseudoscalar $\phi$ and the pion decay constant $f$. $\phi$ corresponds to the
Goldstone boson originating from spontaneous chiral symmetry breaking.
$V=V_\mu dx^\mu$ and $A=A_\mu dx^\mu$ are the one-forms of vector and axial vector flavor gauge fields. We further introduce the left and right one-forms from 
the chiral field and the gauge fields:
\begin{align}
&\af=(\del_\mu U)U^+ dx^\mu=dUU^+ \no
&\bt=U^+dU \no
&A_L=\frac{1}{2}(V+A) \no
&A_R=\frac{1}{2}(V-A) \nonumber
\end{align}
In terms of these fields, the gauged WZW term is given by:
\begin{align}\label{g_wzw}
&\Gm_{WZ}(U,A_L,A_R)= \Gm_{WZ}(U)\no
&+5Ci\int Tr(A_L\af^3+A_R\bt^3)-5C\int Tr[(dA_LA_L+A_LdA_L)\af+(dA_RA_R+A_RdA_R)\bt] \no
&+5C\int Tr[dA_LdUA_RU^+-dA_Rd(U^+)A_LU]+5C\int Tr(A_RU^+A_LU\bt^2-A_LUA_RU^+\af^2) \no
&+\frac{5C}{2}\int Tr[(A_L\af)^2-(A_R\bt)^2]+5Ci\int Tr(A_L^3\af+A_R^3\bt) \no
&+5Ci\int Tr[(dA_RA_R+A_RdA_R)U^+A_LU-(dA_LA_L+A_LdA_L)UA_RU^+] \no
&+5Ci\int Tr(A_LUA_RU^+A_L\af+A_RU^+A_LUA_R\bt) \no
&+5C\int Tr[A_R^3U^+A_LU-A_L^3UA_RU^++\frac{1}{2}(UA_RU^+A_L)^2].
\end{align}
The chiral field is $SU(N)$ valued because in standard QCD, the diagonal $U(1)_A$ symmetry is broken by anomaly. We will promote the chiral field to be $U(N)$ valued as in \cite{KRS}. This can be realized for example in the large $N_c$ limit, where the axial anomaly is suppressed. Under the general $U(N)_L\times U(N)_R$ gauge transformation parameterized by $U(N)$ matrix valued $\eL$ and $\eR$,
\begin{align}
&\dlt U=i\eL U-iU\eR \no
&\dlt A_L=d\eL+i[\eL,A_L] \no
&\dlt A_R=d\eR+i[\eR,A_R] \nonumber,
\end{align}
${\cal L}$ is left invariant, while the WZW term transforms as:
\begin{align}
\dlt\Gm_{WZ}(U,A_L,A_R)=-10Ci\int Tr\bigg[\eL\((dA_L)^2-\frac{i}{2}d(A_L^3)\)-(L\leftrightarrow R)\bigg].
\end{align}

The form of the WZW term in (\ref{g_wzw}) treats the left and right chiral and 
gauge fields in a symmetric fashion, thus is not invariant under the 
either vector or axial vector gauge transformation. In reality, we need to impose invariance under the vector gauge transformation, which requires subtraction of the Bardeen counter term \cite{bardeen}. The Bardeen counter term corresponding to (\ref{g_wzw}) has been worked out in \cite{KRS}:
\begin{align}
\Gm_c=\Gm_{WZ}(U=1,A_L,A_R).
\end{align}
Therefore, the complete WZW term is given by:
\begin{align}\label{f_wzw}
\Gm'_{WZ}(U,A_L,A_R)=\Gm_{WZ}(U,A_L,A_R)-\Gm_{WZ}(1,A_L,A_R).
\end{align}
The gauge transformation of the complete WZW term takes the following form:
\begin{align}\label{gv_wzw}
\dlt\Gm'_{WZ}=30Ci\int Tr(\eR-\eL)\bigg[\frac{1}{4}F_V^2+\frac{1}{12}F_A^2+\frac{i}{6}(F_VA^2+AF_VA+A^2F_V)-\frac{1}{6}A^4\bigg],
\end{align}
with $F_V=dV-\frac{i}{2}(V^2+A^2)$ and $F_A=dA-\frac{i}{2}(VA+AV)$\footnote{There appears to be some minor sign errors in the original paper. We quote the corrected result}.

Now we wish to derive hydrodynamic currents from (\ref{f_S}), (\ref{g_wzw})
and (\ref{f_wzw}). We require the hydrodynamic currents to be gauge invariant (covariant) for abelian (non-abelian) currents. In this section, we will focus on the abelian currents and generalize to the non-abelian currents in the next section. For concreteness and
possible connections to real world situation, we focus on three currents: 
baryonic current $j_B^\mu$, axial baryonic current $j_{B5}^\mu$ and electro-magnetic (EM) current $j_Q^\mu$. We will see that the baryonic current and the EM
current are conserved while the axial baryonic current is not due to anomaly.
We will turn on three external gauge fields: baryonic field $B^\mu$, axial baryonic field $B_5^\mu$ and electro-magnetic $A_{em}^\mu$, all in the Cartan subgroup of $U(N)_L\times U(N)_R$. The first two do not exist in nature. We use them as a way to introduce baryonic and axial baryonic chemical potential. In the end, we put $B^\mu=(\mu,{\vec 0})$ and $B_5^\mu=(\mu_5,{\vec 0})$.

On general grounds, we expect energy-momentum (non)conservation to be one of the hydrodynamic equations:
\begin{align}\label{EM_cons}
\del_\mu T^{\mu\nu}=F^{\nu\lam}_aj_\lam^a.
\end{align}
where $a=B,B_5,Q$ corresponding to baryonic, axial baryonic and EM field respectively. The right hand side (rhs) is a sum over works done by external fields. We will see (\ref{EM_cons}) naturally emerges as a result of the diffeomorphism invariance and gauge invariance. To derive (\ref{EM_cons}), we start from the diffeomorphism invariance of the theory in curved spacetime:
\begin{align}
0=\dlt S\equiv \int d^4x\frac{\dlt S}{\dlt g^{\mu\nu}}\dlt g^{\mu\nu}+\int d^4x\frac{\dlt S}{\dlt U}\dlt U+\int d^4x\frac{\dlt S}{\dlt A_\mu^a}\dlt A_\mu^a.
\end{align}
The diffeomorphism of the metric, chiral and external gauge fields are given by:
\begin{align}
\dlt g^{\mu\nu}=-2\nabla^{(\mu} V^{\nu)},\;\dlt U=V^\mu\del_\mu U,\;\dlt A_\mu^a=V^\nu\del_\nu A_\mu^a+(\del_\mu V^\nu)A_\nu^a.
\end{align}
Note $\frac{\dlt S}{\dlt U}=0$ by equation of motion (EOM) of $U$, we are left with:
\begin{align}
0&=-\int d^4x2\nabla^\mu V^\nu\frac{\dlt S}{\dlt g_{\mu\nu}}+\int d^4x\frac{\dlt S}{\dlt A_\mu^a}(V^\nu\del_\nu A_\mu^a+(\del_\mu V^\nu) A_\nu^a) \no
&=\int d^4x\sqrt{-g}V^\nu\nabla^\mu(\frac{2}{\sqrt{-g}}\frac{\dlt S}{\dlt g^{\mu\nu}})+\int d^4x V^\nu\(\del_\nu A_\mu^a\frac{\dlt S}{\dlt A_\mu^a}-\del_\mu\(\frac{\dlt S}{\dlt A_\mu^a}A_\nu^{a}\)\).
\end{align}
Since $V^\nu$ is arbitrary, we obtain:
\begin{align}\label{EM_cons_C}
\nabla^\mu T_{\mu\nu}=F^{a}_{\nu\mu} J^{\mu,a}-A_\nu^{a}\del_\mu J^{\mu,a},
\end{align}
with the gauge invariant energy-momentum tensor $T_{\mu\nu}=-\frac{2}{\sqrt{-g}}\frac{\dlt S}{\dlt g^{\mu\nu}}$. Note that the current $J^{\mu,a}$ we obtained from the functional derivative is the $consistent$ current. It is neither gauge invariant nor conserved. In order to formulate a hydrodynamic theory, we wish to have a gauge invariant current \cite{oz}. This can be done by expressing the non-conservation of $J^{\mu,a}$ as a correction to $J^{\mu,a}$ \cite{nicolis}:
$A^{\nu,a}\del_\mu\(\frac{\dlt S}{\dlt A_\mu^a}\)=-F^{\nu,a}_\mu\Delta J^{\mu,a}$,
so that the total current $j^{\mu,a}=J^{\mu,a}+\Delta J^{\mu,a}$ is gauge invariant.

To calculate $\del_\mu\(\frac{\dlt S}{\dlt A_\mu^a}\)$, we consider the gauge
transformation of the action:
\begin{align}
\dlt S=\int d^4x\(\frac{\dlt S}{\dlt U}\dlt U+\frac{\dlt S}{\dlt A_\mu}\dlt A_\mu\).
\end{align}
We can again drop the first term in the bracket by EOM of $U$. If the gauge symmetry is not spoiled by anomaly, $\dlt S=0$. For Cartan subgroup, we have
in general $\dlt A_\mu=\del_\mu\eps$. Therefore we conclude that the current 
from functional derivative is conserved:
\begin{align}
0=\int d^4x\frac{\dlt S}{\dlt A_\mu}\del_\mu\eps \Rightarrow \del_\mu\(\frac{\dlt S}{\dlt A_\mu}\)=0.
\end{align}
If the gauge group is anomalous, $\dlt S$ is nonvanishing, which should be used to obtain the nonvanishing $\del_\mu\(\frac{\dlt S}{\dlt A_\mu}\)$. We are 
interested in three particular gauge transformations list below, which correspond to EM gauge, baryonic gauge and axial baryonic gauge respectively:
\begin{align}\label{gauges}
&\text{gauge Q:}\;\eL=\eR=\eps,\;  \dlt U=i\eps[Q,U],\;\dlt A_{em}=2d\eps,\;\dlt B=\dlt B_5=0 \\
&\text{gauge B:}\;\eL=\eR=\eps,\; \dlt U=0,\; \dlt B=2d\eps,\;\dlt B_5=\dlt A_{em}=0 \\
&\text{gauge B5:}\;\eL=-\eR=\eps,\; \dlt U=2i\eps U,\; \dlt B_5=2d\eps,\; \dlt A_{em}=\dlt B=0.
\end{align}
They can be easily obtained from (4.17) of \cite{KRS}. Note in the above, all 
$\eps's$ are just numbers instead of matrices.

From (\ref{gv_wzw}), we see $J_B$ and $J_Q$ are obviously conserved, while $J_{B5}$ is not. We calculate its divergence as follows:
\begin{align}\label{db5}
\dlt S=\int J_{B5}\dlt B_5=\int 2J_{B5}d\eps=-2\int \eps dJ_{B5}.
\end{align}
On the other hand, our external gauge fields lead to
\begin{align}
&V=B+\Aq Q,\; A=B_5 \\
&F_V=dB+d\Aq Q,\; F_A=dB_5.
\end{align}
Plugging them into (\ref{gv_wzw}) and compare with (\ref{db5}), we obtain
\begin{align}
dJ_{B5}=\frac{15Ci}{2}\((dB)^2Tr1+2dBd\Aq TrQ+(d\Aq)^2TrQ^2\)+\frac{5Ci}{2}(dB_5)^2Tr1.
\end{align}

Now we want to obtain explicit expressions for the $consistent$ currents $J_B$, $J_{B5}$ and $J_Q$. We will focus on the contributions from the WZW term. The contributions from the original chiral Lagrangian can always be included later. It is easier to start with the non-abelian(matrix valued) left-right current as follows \cite{BPR}:
\begin{align}\label{J_L}
&J_L\equiv\frac{\dlt \Gm_{WZ}(U,A_L,A_R)}{\dlt A_L} \no
&=5Ci\af^3-5C(2dA_L\af-A_L d\af+2\af dA_L-d\af A_L)+5C\af A_L\af+5Ci(-A_L\af A_L) \no
&+5C(-dUdA_RU^++dUA_RdU^++UdA_RdU^+)+5C(-U\bt^2 A_RU^+-UA_RU^+\af^2) \no
&+5Ci\bigg[-U(dA_RA_R+A_RdA_R)U^+-2dA_LUA_RU^++A_LdUA_RU^++A_LUdA_RU^+ \no
&-A_LUA_RdU^+-2UA_RU^+dA_L+dUA_RU^+A_L+UdA_RU^+A_L-UA_RdU^+A_L\bigg] \no
&+5Ci(UA_RU^+A_L\af+\af A_LUA_RU^+-UA_R\bt A_RU^+) \no
&+5C(A_LUA_RU^+A_L-UA_RU^+A_LUA_RU^+),
\end{align}
\begin{align}\label{J_R}
&J_R\equiv\frac{\dlt \Gm_{WZ}(U,A_L,A_R)}{\dlt A_R} \no
&=5Ci\bt^3-5C(2dA_R\bt-A_Rd\bt+2\bt dA_R-d\bt A_R)+5C(-\bt A_R\bt)+5Ci(-A_R\bt A_R) \no
&+5C(-U^+dA_LdU+dU^+dA_LU-dU^+A_LdU)+5C(U^+A_LU\bt^2+U^+\af^2A_LU) \no
&+5Ci\bigg[2dA_RU^+A_LU-A_RdU^+A_LU-A_RU^+dA_LU+A_RU^+A_LdU+2U^+A_LUdA_R \no
&-dU^+A_LUA_R-U^+dA_LUA_R+U^+A_LdUA_R+U^+(dA_LA_L+A_LdA_L)U\bigg] \no
&+5Ci(-U^+A_L\af A_LU+U^+A_LUA_R\bt+\bt A_RU^+A_LU) \no
&+5C(-A_RU^+A_LUA_R+U^+A_LUA_RU^+A_LU).
\end{align}
The abelian currents we are interested in are simply given by:
\begin{align}\label{three_j}
&J_B=\frac{1}{2}Tr(J_L+J_R)-(U\rightarrow 1) \\
&J_{B5}=\frac{1}{2}Tr(J_L-J_R)-(U\rightarrow 1) \\
&J_Q=\frac{1}{2}Tr Q(J_L+J_R))-(U\rightarrow 1).
\end{align}
Plugging $A_L=\frac{B+B_5+\Aq Q}{2}$ and $A_R=\frac{B-B_5+\Aq Q}{2}$ into 
(\ref{J_L}), (\ref{J_R}) and (\ref{three_j}), we obtain after tedious algebra:
\begin{align}\label{J_exp}
&J_B=\frac{5Ci}{2}Tr(\af^3+\bt^3)-\frac{15C}{2}(dB Tr(\af+\bt)+d\Aq TrQ(\af+\bt)-\Aq TrQ(\af^2-\bt^2)) \no
&J_{B5}=-\frac{5C}{2}(dB_5Tr(\af+\bt)+d\Aq TrQ(\af-\bt))+\frac{5Ci}{2}\Aq d\Aq(TrQ^2-TrU^+QUQ) \no
&J_Q=\frac{5Ci}{2}TrQ(\af^3+\bt^3)+\frac{5C}{2}\bigg[-3dB TrQ(\af+\bt)-dB_5 TrQ(\af-\bt)+B_5TrQ(\af^2+\bt^2) \no
&-d\Aq(2TrQ^2(\af+\bt)+TrQdUQU^+-TrQUQdU^+)+\Aq(TrQ^2(\af^2-\bt^2)-TrQdUQdU^+)\bigg] \no
&\frac{5Ci}{2}(dB_5\Aq+2d\Aq B_5)(TrQUQU^+-TrQ^2)-\frac{5Ci}{2}B_5\Aq Tr(QdUQU^++QdU^+QU).
\end{align}
We can check explicitly that all three currents are invariant under $B$ and $Q$
gauges, but none is invariant under $B_5$ gauge. The verification of the gauge
invariance is lengthy. In the next section, we will see it is much more convenient to use the covariant variables.

Now we want to correct (\ref{J_exp}) so that they are $B_5$ gauge invariant
also. We first write (\ref{EM_cons_C}) out explicitly:
\begin{align}
\del^\mu T_{\mu\lam}=F^{B}_{\lam\mu} J_B^\mu+F^Q_{\lam\mu} J_Q^\mu+F^{B5}_{\lam\mu} J_{B5}^\mu-B_{5\lam}\del_\mu J_{B5}^\mu.
\end{align}
We wish to express $B_5dJ_{B5}$ as a sum of $F*J$ terms. The former in terms
of components of gauge fields, is given by:
\begin{align}
&B_{5\lam}\del_\mu J_{B5}^\mu \no
=&B_{5\lam}\bigg[\frac{15Ci}{2}\eps^{\mu\nu\af\bt}(\del_\mu B_\nu\del_\af B_\bt Tr1+2\del_\mu B_\nu \del_\af \Aq{}_{,\bt} TrQ+\del_\mu\Aq{}_{,\nu}\del_\af\Aq{}_{,\bt} TrQ^2) \no
&+\frac{5Ci}{2}\eps^{\mu\nu\af\bt}\del_\mu B_{5\nu}\del_\af B_{5\bt}Tr1 \bigg].
\end{align}
Indeed, it can be expressed as
\begin{align}\label{identity}
B_{5,\lam}\del_\mu J_{B5}^\mu=F^B_{\lam\mu}\Delta J_B^\mu+F^{B5}_{\lam\mu}\Delta J_{B5}^\mu+F^{Q}_{\lam\mu}\Delta J_Q^\mu,
\end{align}
where
\begin{align}\label{delta_J}
&\Delta J_B^\mu=15Ci\eps^{\mu\nu\af\bt}\del_\nu B_\af B_{5\bt}Tr1+15Ci\eps^{\mu\nu\af\bt}\del_\nu\Aq{}_{,\af} B_{5\bt}TrQ \\
&\Delta J_{B5}^\mu=5Ci\eps^{\mu\nu\af\bt}\del_\nu B_{5\af}B_{5\bt}Tr1 \\
&\Delta J_Q^\mu=15Ci\eps^{\mu\nu\af\bt}\del_\nu B_\af B_{5\bt}TrQ+15Ci\eps^{\mu\nu\af\bt}\del_\nu\Aq{}_{,\af} B_{5\bt}TrQ^2.
\end{align}
(\ref{identity}) can be proved by repeated use of the following identity:
\begin{align}
(\del_\lam X_\mu-\del_\mu X_\lam)\eps^{\mu\nu\af\bt}\del_\nu Y_\af\ Z_\bt+(\del_\lam Y_\mu-\del_\mu Y_\lam)\eps^{\mu\nu\af\bt}\del_\nu X_\af Z_\bt+Z_\lam\eps^{\mu\nu\af\bt}\del_\mu X_\nu\del_\af Y_\bt=0.
\end{align}
As a result, we obtain our final results of the gauge invariant currents:
\begin{align}\label{3jc}
j_B=J_B+\Delta J_B,\;j_B=J_{B5}+\Delta J_{B5},\;j_Q=J_Q+\Delta J_Q.
\end{align}
We can check explicitly that the currents $j_B$, $j_{B5}$ and $j_Q$ are invariant under all three gauge transformation.

\section{Extension to non-abelian currents}

We also wish to obtain all non-abelian currents, which are (\ref{J_L}) and (\ref{J_R}) with properly chosen corrections to restore gauge covariance. We again turn on only the electro-magnetic field and the fictitious baryonic and axial baryonic fields. We do not turn on isospin chemical potentials, because in the chiral limit we are working at, a finite isospin chemical potential will lead to Bose-Einstein condensation of the associated Goldstone boson \cite{isospin}, thus further reducing the number of Goldstone bosons.

It is convenient to work with the covariant quantities defined as follows:
\begin{align}\label{cov}
&D U=dU-iA_LU+iUA_R,\;D U^+=dU^+-iA_RU^++iU^+A_L \no
&\baf=D UU^+,\; \bbt=U^+D U. 
\end{align}
In terms of the covariant quantities, the consistent currents (\ref{J_L}) and (\ref{J_R}) take the following compact forms:
\begin{align}\label{J_cov}
J_L&=5Ci\baf^3-5C\bigg[4idA_LA_L+2(dA_L\baf+\baf dA_L)\bigg] \no
&+5C(-D UdA_RU^++UdA_RD U^+) \no
J_R&=5Ci\bbt^3-5C\bigg[-4idA_RA_R+2(dA_R\bbt+\bbt dA_R)\bigg] \no
&+5C(-U^+dA_LD U+D U^+dA_LU).
\end{align}
In arriving at (\ref{J_cov}), we have used $A_L^2=A_R^2=0$ and the commutative relation among $A_L$, $A_R$, $dA_L$ and $dA_R$. Subtracting the Bardeen counter terms, we obtain:
\begin{align}\label{Jc_cov}
J_L^c&\equiv J_L-J_L(U\to1) \no
&=5Ci\baf^3-10C(dA_L\baf+\baf dA_L)+5C(-D UdA_RU^++UdA_RD U^+) \no
&-5C(4idA_LA_L-4idA_LA_R-2idA_RA_R+2idA_RA_L) \\
J_R^c&\equiv J_R-J_R(U\to1) \no
&=5Ci\bbt^3-10C(dA_R\bbt+\bbt dA_R)+5C(-U^+dA_LD U+D U^+dA_LU) \no
&-5C(-4idA_RA_R+4idA_RA_L+2idA_LA_L-2idA_LA_R).
\end{align}

It is straightforward to work out the gauge variations of the left and right currents. Starting with the gauge variations of the covariant quantities:
\begin{align}\label{gauges_cov}
\text{gauge B:}\;&\dlt U=0,\;\dlt A_L=\dlt A_R=d\eps,\; \dlt D U=\dlt D U^+=0,\;
\dlt\baf=\dlt\bbt=0 ,\\
\text{gauge B5:}\;&\dlt U=2i\eps U,\;\dlt A_L=-\dlt A_R=d\eps,\; \dlt D U=2i\eps D U,\;\dlt D U^+=-2i\eps D U^+,\no
&\dlt\baf=\dlt\bbt=0, \\
\text{gauge Q:}\;&\dlt U=i\eps[Q,U],\;\dlt A_L=\dlt A_R=d\eps Q,\;\dlt D U=i\eps[Q,D U],\;\dlt D U^+=i\eps[Q,D U^+],\no
&\dlt\baf=i\eps[Q,\baf],\;\dlt\bbt=i\eps[Q,\bbt] ,
\end{align}
we readily obtain the gauge variations of the $consistent$ currents as follows:
\begin{align}\label{gauge_QBB5}
\text{Q:}&\; \dlt J_L^c=i\eps[Q,J_L^c],\; J_R^c=i\eps[Q,J_R^c] \\
\text{B:}&\; \dlt J_L^c=\dlt J_R^c=0 \\
\text{B5:}&\; \dlt J_L^c=-5C(8idA_Ld\eps+4idA_R d\eps),\;
\dlt J_R^c=-5C(4idA_L d\eps+8idA_Rd\eps) .
\end{align}
Similar to the case of abelian currents in the previous section, the non-abelian currents are Q-gauge covariant and B-gauge invariant, but are not B5 gauge invariant. We need corrections to the currents to restore the gauge invariance. We note that the corrections are simply given by the $U$ independent terms:
\begin{align}\label{correction}
&\Delta J_L^c=5C(4idA_LA_L-4idA_LA_R+2idA_RA_L-2idA_RA_R) \no
&\Delta J_R^c=5C(4idA_RA_L-4idA_RA_R+2idA_LA_L-2idA_LA_R).
\end{align}
Our final results for the $covariant$ currents are given by:
\begin{align}\label{JVA}
&j_L^c=J_L^c+\Delta J_L^c \no
&=5Ci\baf^3-10C(dA_L\baf+\baf dA_L)+5C(-UA_RU^+A_LUA_RU^+-D UdA_RU^++UdA_RD U^+) \\
&j_R^c=J_R^c+\Delta J_R^c \no
&=5Ci\bbt^3-10C(dA_R\bbt+\bbt dA_R)+5C(-U^+dA_LD U+D U^+dA_LU+U^+A_LUA_RU^+A_LU) .
\end{align}
We can verify that $\frac{1}{2}Tr(j_L^c+j_R^c)$, $\frac{1}{2}Tr(j_L^c-j_R^c)$ and $\frac{1}{2}TrQ(j_L^c+j_R^c)$ reproduce the abelian currents $j_B$, $j_{B5}$ and $j_Q$ in (\ref{3jc}) respectively.

The non(conservation) of the covariant currents $j_V^c$ and $j_A^c$ can be obtained from (\ref{gv_wzw}). The most general gauge variation is given by:
\begin{align}
\dlt S&=\int Tr\bigg[\frac{\dlt S}{\dlt A_L}(d\eL+i[\eL,A_L])+\frac{\dlt S}{\dlt A_R}(d\eR+i[\eR,A_R])\bigg] \no
&=\int Tr\bigg[\eL(-dJ_L^c+i[A_L,J_L^c])+\eR(-dJ_R+i[A_R,J_R])\bigg] \no
&=\int Tr(\eL D_LJ_L^c+\eR D_RJ_R),
\end{align}
where $D_L=d-i[A_L,\,]$ and $D_R=d-i[A_R,\,]$. With our choice of the external gauge fields, the left-right covariant derivatives simplify to $D_L=D_R=d-\frac{i\Aq}{2}[Q,\,]\equiv \Dq$. Comparing with (\ref{gv_wzw}), we readily obtain
\begin{align}
\Dq J_L^c=30Ci(\frac{1}{4}F_V^2+\frac{1}{12}F_A^2),\; \Dq J_R^c=-30Ci(\frac{1}{4}F_V^2+\frac{1}{12}F_A^2).
\end{align}
Combined with the corrections to the currents (\ref{correction}), we have
\begin{align}
\Dq j_L^c=30Ci(dA_L)^2,\; \Dq j_R^c=-30Ci(dA_R)^2.
\end{align}

\section{A hydrodynamic description of the chiral superfluid}

\subsection{non-abelian hydrodynamics without anomaly}
In this section, we attempt to formulate a hydrodynamic description of the chiral superfluid. This is a zero temperature version of \cite{son}. The new ingredients are the effect of anomaly and external electromagnetic field and chemical potentials. We start with the action without the WZW term. The stress tensor and currents can be easily obtained from functional derivatives\footnote{The non-anomalous currents are automatically covariant and conserved.}:
\begin{align}\label{LO}
T_{\mu\nu}&=-\frac{f^2}{4}\(Tr D_\mu U D_\nu U^+-\frac{1}{2}g_{\mu\nu}Tr D_\lam U D^\lam U^+\) \no
&=-\frac{f^2}{4}\(Tr\(iD_\mu UU^+\,iD_\nu UU^+\)-\frac{1}{2}g_{\mu\nu}Tr\(iD_\lam UU^+\,iD^{\lam} UU^+\)\) \\
J_{0L}^\mu&=\frac{f^2}{4}iD^\mu UU^+ \\
J_{0R}^\mu&=-\frac{f^2}{4}iU^+D^\mu U,
\end{align}
where we have used an index $0$ in the currents to denote the contributions from the non-anomalous part of the action.
The dynamical equations are simply
\begin{subequations}
\begin{align}
&\del_\mu T^{\mu\nu}=F_Q^{\nu\mu}J_{0\mu}^Q+F^{\nu\mu}_{B5}J_{0\mu}^{B5} \label{LO_eq} \\
&\Dq_\mu J_{0L}^\mu=\Dq_\mu J_{0R}^\mu=0 \label{LO_jeq}, 
\end{align}
\end{subequations}
with $J_{0Q}^\mu=\frac{1}{2}TrQ\(J_{0L}^\mu+J_{0R}^\mu\),\;J_{0,B5}^\mu=\frac{1}{2}Tr\(J_{0L}^\mu-J_{0R}^\mu\)$ and $F_Q=dA^{em},\;F_{B5}=dB_5$. Note that the baryonic gauge field $B$ is absent in (\ref{LO_eq}) and (\ref{LO_jeq}). The axial gauge field $B_5$ is used to model the axial chemical potential $B_5^\mu=(\mu_5,{\vec 0})$. Semi-classically it originates from fluctuations of instantons in vacuum. We separate its contribution to the covariant derivative $D^\mu UU^+=D_Q^\mu UU^+-iB_5^\mu$ due to its special role.

In \cite{son}, the phase of the condensate $U$ was chosen as the dynamical variables, we instead use $i\Dq_\mu UU^+$ and $iU^+\Dq_\mu U$, which has an interpretation the non-abelian generalization of superfluid velocity. To justify this interpretation, we note that $i\Dq_\mu UU^+$ and $iU^+\Dq_\mu U$ are hermitian (In abelian theory, the hermiticity condition will be replaced by the reality condition for the superfluid velocity). Their decomposition on the basis of $U(N)$ generators all have real entries. The hermiticity can be seen by noting the following identity
\begin{align}
\Dq_\mu UU^++U\Dq_\mu U^+=0.
\end{align}
Therefore, we define the non-abelian superfluid velocity as:
\begin{align}\label{super_v}
i\Dq_\mu UU^+=\xiL_\mu=\xi^{La}_\mu T^a,\;-iU^+\Dq_\mu U=\xiR_\mu=\xi^{Ra}_\mu T^a,
\end{align}
with $T^a$ being the $U(N)$ generators. In terms of the new variables, the constitutive equations become:
\begin{align}\label{const}
T^{\mu\nu}&=\frac{f^2}{4}\(Tr(\xiL{}^\mu+B_5^\mu)(\xiL{}^\nu+B_5^\nu)-\frac{1}{2}g^{\mu\nu}Tr(\xiL{}^{\lam}+B_5^\lam)(\xiL_{\lam}+B_{5\lam})\) \no
J_{0L}^\mu&=\frac{f^2}{4}\(\xiL{}^\mu+B_5^\mu\) \no
J_{0R}^\mu&=\frac{f^2}{4}\(\xiR{}^\mu-B_5^\mu\) \no
J_{0Q}^{\mu}&=\frac{f^2}{8}TrQ\(\xiL{}^\mu+\xiR{}^\mu\).
\end{align}
Treating $B_5^\mu$ as non-dynamical field, we can derive the conservation of energy-momentum tensor (\ref{LO_eq}) from the conservation of left and right currents (\ref{LO_jeq}). Therefore (\ref{LO_eq}) is redundant thus can be discarded from the dynamical equations. We could make $B_5^\mu$ dynamical by writing $B_5^\mu=\mu_5u^\mu$ and interpreting $u^\mu$ as a normal component in the hydrodynamic theory. Note that at zero temperature $B_5^\mu$ originates from the fluctuation of instantons in vacuum. A dynamical $B_5^\mu$ would mean we include the dynamics of the gluons. We will not do so in this paper, but only treat $B_5^\mu$ as an external field.

The defined superfluid velocities allow us to have a compact form for the constitutive equation. The price to pay is an increased number of unknowns. We have two $4-$component superfluid velocities, in total $8$ unknowns, but only $2$ dynamical equations. The system of equations does not close. Note that components of $\xiL_\mu$ and $\xiR_\mu$ are not independent. It is not difficult to prove the following identities:
\begin{align}\label{xi_com}
&\Dq_\nu\xiL_\mu-\Dq_\mu\xiL_\nu=i[\xiL_\mu,\xiL_\nu]+[Q,U]U^+\frac{F^Q_{\nu\mu}}{2} \no
&\Dq_\nu\xiR_\mu-\Dq_\mu\xiR_\nu=i[\xiR_\mu,\xiR_\nu]-U^+[Q,U]\frac{F^Q_{\nu\mu}}{2}.
\end{align}
We can close the system by taking the $\mu,\nu=0,i$ component of (\ref{xi_com}),i.e.
\begin{align}\label{xi_comd}
&\Dq_t\xiL_i-\Dq_i\xiL_t=i[\xiL_i,\xiL_t]-(Q-UQU^+)\frac{E_i}{2} \no
&\Dq_t\xiR_i-\Dq_i\xiR_t=i[\xiR_i,\xiR_t]+(U^+QU-Q)\frac{E_i}{2},
\end{align}
which provide dynamical equations for $\xiL_i$ and $\xiR_i$. We have used $F^Q_{i0}=E_i$. Unlike the external magnetic field, which affects the dynamics, but not pump energy into the system, the external electric can possibly heat up the system and upset our zero temperature description of the system. However, since the EM field remains entirely arbitrary, we can always have an electric field that is orthogonal to the currents and does no work to the system.

Note the appearance of $UQU^+$ and $U^+QU$ in (\ref{xi_comd}) mean we have not complete the equations of motion. Defining the hermitian variables $\xL=UQU^+$ and $\xR=U^+QU$, we easily find the equations satisfied by $\xL$ and $\xR$:
\begin{align}\label{xLR}
&\Dq_\mu\xL=-i[\xiL_\mu,\xL] \no
&\Dq_\mu\xR=-i[\xiR_\mu,\xR].
\end{align}
As in the case of $\xiL{}^\mu$ and $\xiR{}^\mu$, the time components of (\ref{xLR}) provide the dynamical equations for $\xL$ and $\xR$.
To summarize, we have found the following hydrodynamic equations:
\begin{align}\label{LO_extend}
&\Dq_\mu J^\mu_{0L}=\Dq_\mu J^\mu_{0R}=0 \no
&\Dq_t\xiL_i-\Dq_i\xiL_t=i[\xiL_i,\xiL_t]-(Q-\xL)\frac{E_i}{2} \no
&\Dq_t\xiR_i-\Dq_i\xiR_t=i[\xiR_i,\xiR_t]+(\xR-Q)\frac{E_i}{2} \no
&\Dq_t\xL=-i[\xiL_t,\xL] \no
&\Dq_t\xR=-i[\xiR_t,\xR].
\end{align}
The first line provides the dynamical equations for $\xiL_t$ and $\xiR_t$. The rest of the equations provide dynamical equations for $\xiL_i$, $\xiR_i$, $\xL$ and $\xR$ respectively. The additional equations given by (\ref{xi_com}) and (\ref{xLR}) are non-dynamical. They are just constraint equations, which should be satisfied by the initial conditions. Once satisfied by the initial condition, the constraint equations will continue to hold as the system evolves.

\subsection{non-abelian hydrodynamics with anomaly}

Now we are ready to consider the effect of the WZW term on the hydrodynamics. The WZW term does not have correction to the LO of the stress tensor, which can be easily verified. It does induce corrections to the non-abelian currents. We have worked these out in the previous section. We quote the results as follows:
\begin{align}
&j_L=5Ci\baf^3-10C(dA_L\baf+\baf dA_L)+5C(-\baf UdA_RU^+-UdA_RU^+\baf) \no
&j_R=5Ci\bbt^3-10C(dA_R\bbt+\bbt dA_R)+5C(-U^+dA_LU\bbt-\bbt U^+dA_LU)
\end{align}
In terms of explicit components, we have
\begin{align}\label{wzw_J}
j_L^\mu&=-5C\eps^{\mu\nu\rho\sig}\(\xiL_\nu\xiL_\rho\xiL_\sig+\xiL_\nu\xiL_\rho B_{5\sig}\) \no
&+\frac{5Ci}{2}\eps^{\mu\nu\rho\sig}\bigg[\del_\nu A^{em}_\rho(2Q+\xL)(\xiL_\sig+B_{5\sig})+(\xiL_\sig+B_{5\sig})(2Q+\xL)\del_\nu A^{em}_\rho\bigg] \no
j_R^\mu&=5C\eps^{\mu\nu\lam\rho}\(\xiR_\nu\xiR_\rho\xiR_\sig-\xiR_\nu\xiR_\rho B_{5\sig}\) \no
&+\frac{5Ci}{2}\eps^{\mu\nu\rho\sig}\bigg[\del_\nu A^{em}_\rho(2Q+\xR)(-\xiR_\sig+B_{5\sig})+(-\xiR_\sig+B_{5\sig})(2Q+\xR)\del_\nu A^{em}_\rho\bigg] .
\end{align}
where we have used the abelian nature of $B_5^\mu$ to simplify the expressions.
We can use (\ref{xi_com}) to rewrite (\ref{wzw_J}) into a more suggestive form:
\begin{align}\label{wzw_J2}
j_L^\mu&=-\frac{5Ci}{2}\eps^{\mu\nu\rho\sig}\(\Dq_\nu\xiL_\rho(\xiL_\sig+B_{5\sig})+(\xiL_\sig+B_{5\sig})\Dq_\nu\xiL_\rho\) \no
&+\frac{5Ci}{2}\eps^{\mu\nu\rho\sig}\bigg[\del_\nu A^{em}_\rho(\frac{5Q}{2}+\frac{\xL}{2})(\xiL_\sig+B_{5\sig})+(\xiL_\sig+B_{5\sig})(\frac{5Q}{2}+\frac{\xL}{2})\del_\nu A^{em}_\rho\bigg] \no
j_R^\mu&=\frac{5Ci}{2}\eps^{\mu\nu\rho\sig}\(\Dq_\nu\xiR_\rho(\xiR_\sig-B_{5\sig})+(\xiR_\sig-B_{5\sig})\Dq_\nu\xiR_\rho\) \no
&-\frac{5Ci}{2}\eps^{\mu\nu\rho\sig}\bigg[\del_\nu A^{em}_\rho(\frac{5Q}{2}+\frac{\xR}{2})(\xiL_\sig-B_{5\sig})+(\xiR_\sig-B_{5\sig})(\frac{5Q}{2}+\frac{\xR}{2})\del_\nu A^{em}_\rho\bigg].
\end{align}

Due to the effect of the WZW term, the hydrodynamic equations are slightly modified as follows:
\begin{align}\label{wzw_hydro}
&\Dq_\mu (J^\mu_{0L}+j^\mu_L)=30Ci(dA_L)^2=\frac{15Ci}{8}\eps^{\mu\nu\rho\sig}F^Q_{\mu\nu}F^Q_{\rho\sig}Q^2 \no
&\Dq_\mu (J^\mu_{0R}+j^\mu_R)=-30Ci(dA_R)^2=-\frac{15Ci}{8}\eps^{\mu\nu\rho\sig}F^Q_{\mu\nu}F^Q_{\rho\sig}Q^2 \no
&\Dq_t\xiL_i-\Dq_i\xiL_t=i[\xiL_i,\xiL_t]-(Q-\xL)\frac{E_i}{2} \no
&\Dq_t\xiR_i-\Dq_i\xiR_t=i[\xiR_i,\xiR_t]+(\xR-Q)\frac{E_i}{2} \no
&\Dq_t\xL=-i[\xiL_t,\xL] \no
&\Dq_t\xR=-i[\xiR_t,\xR].
\end{align}
Note that the external baryonic field $B^\mu$ or the baryon chemical potential does not appear in the hydrodynamic variables and equations. However, a nonvanishing baryon chemical potential can change the ground state of the theory, thus implicitly entering the expressions for the currents. We will see such an example in the next section.

The baryonic and axial baryonic currents defined in (\ref{three_j}) are of particular interest. The relevant contributions are given by:
\begin{align}\label{Jj_com}
&Tr\(J_{0L}^\mu+J_{0R}^\mu\)=0 \no
&Tr\(J_{0L}^\mu-J_{0R}^\mu\)=\frac{f^2}{4}\(Tr(\xiL{}^\mu-\xiR{}^\mu)+2NB_5^\mu\) \no
&Tr\(j_L^\mu+j_R^\mu\)=-5Ci\eps^{\mu\nu\rho\sig}Tr\(\Dq_\nu\xiL_\rho\xiL_\sig-\Dq_\nu\xiR_\rho\xiR_\sig\) \no
&+5Ci\eps^{\mu\nu\rho\sig}\bigg[3\del_\nu A^{em}_\rho TrQ(\xiL_\sig-\xiR_\sig)+6\del_\nu A^{em}_\rho TrQB_{5\sig}\bigg] \no
&Tr\(j_L^\mu-j_R^\mu\)=10Ci\eps^{\mu\nu\rho\sig}\del_\nu A^{em}_\rho TrQ(\xiL_\sig+\xiR_\sig).
\end{align}
The anomalous baryonic current can be written alternatively as:
\begin{align}
&Tr\(j_L^\mu+j_R^\mu\)=-10C\eps^{\mu\nu\rho\sig}Tr\xiL_\nu\xiL_\rho\xiL_\sig \no
&+5Ci\eps^{\mu\nu\rho\sig}\bigg[3\del_\nu A^{em}_\rho TrQ(\xiL_\sig-\xiR_\sig)+6\del_\nu A^{em}_\rho TrQB_{5\sig}\bigg],
\end{align}
which agrees with the previous results \cite{BPR,domain} upon setting $B^\mu_5=0$.
It is illuminating to separate the temporal and spatial components of the currents induced by WZW term:
\begin{align}\label{oi_com}
&Tr(j_L^0+j_R^0)=-5Ci\eps^{ijk}Tr\(\Dq_i\xiL_j\xiL_k-\Dq_i\xiR_j\xiR_k\)+15CiB_iTrQ(\xiL_i-\xiR_i) \no
&Tr(j_L^i+j_R^i)=-5CiTr(\Omg^L_i-\Omg^R_i) \no
&+15Ci\bigg[\eps^{ijk}E_jTrQ(\xiL_k-\xiR_k)-B_iTrQ(\xiL_0-\xiR_0)+2\mu_5B_iTrQ\bigg] \no
&Tr(j_L^0-j_R^0)=10CiB_iTrQ(\xiL_i+\xiR_i) \no
&Tr(j_L^i-j_R^i)=10Ci\(\eps^{ijk}E_jTrQ(\xiL_k+\xiR_k)-B_iTrQ(\xiL_0+\xiR_0)\),
\end{align}
where we have used $B^\mu_5=(\mu_5,{\vec 0})$. We have also defined $\Omg^L_i=\eps^{i\nu\rho\sig}TrD_\nu\xiL_\rho\xiL_\sig$ and $\Omg^R_i=\eps^{i\nu\rho\sig}TrD_\nu\xiR_\rho\xiR_\sig$, which are analogs of vorticity in non-abelian superfluid, thus giving rise to the CVE. They are vanishing for abelian superfluid. This can be easily shown by using (\ref{xi_com}) and the commutativity of abelian superfluid velocities. $E_i$ and $B_i$ in (\ref{oi_com}) are external electric and magnetic field. The latter is not to be confused with components of $B_5$, which originates from the fluctuation of instantons. The terms proportional to $E_i$ and $B_i$ can be interpreted as the analogs of the CEE and CME. Note that CVE and CME have already been found in \cite{lublin}. The identification of CEE is new in the current analysis. From various terms in (\ref{oi_com}), we observe that there is CVE in the baryonic current and CEE and CME are present in both baryonic and axial baryonic currents.

\section{Application to the ground state}

As an example, we calculate the currents in the ground state of QCD at finite baryonic or axial chemical potential in the presence of a constant external magnetic field. This will be done by solving the hydrodynamic equations (\ref{wzw_hydro}) for the superfluid velocities, the results of which will be used to obtain the baryonic and axial currents from (\ref{Jj_com}).

We focus on one-flavor case in which we have the diagonal $\eta'$ meson only. The calculation simplifies significantly as all the non-abelian structures drop out. It is convenient to use the $\baf$ and $\bbt$ variables in solving the hydrodynamic equations (\ref{wzw_hydro}). Since the field $U$ is just a $U(1)$ phase: $U=e^{i\varphi}$, we obviously have $\baf=\bbt$, $\xL=\xR=Q$, therefore $\xL$ and $\xR$ do not enter the hydrodynamic equations. (\ref{wzw_hydro}) reduces to the following:
\begin{align}
&\del_\mu(J_L^\mu+j_L^\mu)=0 \no
&\del_\mu(J_R^\mu+j_R^\mu)=0 , \\
&\text{where}\no
&J_L=i\frac{f^2}{4}\baf\;, J_R=-i\frac{f^2}{4}\baf \no
&j_L=-15C dA^{em}\baf\; ,j_R=-15C dA^{em}\baf.
\end{align}
We proceed further by considering a plane wave ansatz: $U=e^{ik_\mu x^\mu}$, thus we have $\baf^\mu=i(k^\mu-B_5^\mu)$, with $B_5^\mu=(\mu_5,0,0,0)$. The constant magnetic field is applied along the $z$-axis: ${\vec B}=(0,0,B)$. It is easy to see all the equations are trivially satisfied. To determine the momentum $k^\mu$, we need to minimize the free energy $H-\mu N_B$ and $H-\mu_5N_{B5}$ for finite $\mu$ and $\mu_5$ respectively \cite{thompson}. Let us consider the case $\mu\ne0,\,\mu_5=0$ first:
\begin{align}
H-\mu N_B=\int d^3x\(\frac{f^2}{8}(k_0^2+k_1^2+k_2^2+k_3^2)-15Ci\mu Bk_3\).
\end{align}
It is obviously minimized at $k_\mu=(0,0,0,-\frac{60Ci\mu B}{f^2})$. The resulting nonzero components of the baryonic and axial currents are given by:
\begin{align}\label{CME}
J_B^0=\frac{(30Ci)^2\mu B^2}{f^2},\;J_{B5}^3=15Ci\mu B.
\end{align}
Similarly for the case $\mu=0,\,\mu_5\ne0$, the free energy is given by
\begin{align}
H-\mu_5 N_{B5}=\int d^3x\(\frac{f^2}{8}((k_0+\mu_5)^2+k_1^2+k_2^2+k_3^2)-\frac{f^2}{4}\mu_5(k_0+\mu_5)\).
\end{align}
It is minimized at $k_\mu=(0,0,0,0)$, with the following nonzero components of the baryonic and axial currents:
\begin{align}\label{CSE}
J_B^3=15Ci\mu_5B,\;J_{B5}^0=\frac{f^2\mu_5}{4}.
\end{align}
We identify in (\ref{CME}) and (\ref{CSE}) the CME and chiral separation \cite{topology,GMS} in confined phase. Also a finite baryonic and axial chemical potential induces corresponding charge density respectively, although they differ in the dependences on the external magnetic field.

\section{Conclusion}

Following the idea of \cite{lublin}, we have derived, at the mean field level, a non-abelian superfluid hydrodynamics from the chiral Lagrangian and WZW term. It is a zero temperature hydrodynamic theory, with the dynamical equations involving the non-abelian currents only. The constitutive equations for the currents contain contributions from the chiral Lagrangian and the WZW term. We identify superfluid analogs of CVE, CME and CEE in the contributions from the WZW term. The hydrodynamic equations can be useful in the study of the low energy dynamics of QCD. As an example, we solved the ground state of the hydrodynamic theory at finite baryonic or axial chemical potential in the presence of a constant magnetic field. We obtained the anomaly induced contributions to the baryonic and axial currents, which can be viewed as CME and chiral separation in the confined phase.

Throughout the paper, we have been restricting ourselves to topologically trivial configuration of the chiral field. It has been known that topologically non-trivial configuration such as vortex and domain wall allows for more terms in the effective Lagrangian \cite{topology,domain}. It is interesting to investigate the anomaly induced effect with a relaxed topological condition. 

It is also interesting to consider the effect of mass term and isospin chemical potential for the chiral field, which can lead to the Bose-Einstein condensation of the associated Goldstone boson. It is curious to see how the interplay of chiral condensate and BEC will modify the hydrodynamic description.

\noindent{\large \bf Acknowledgments} \vskip .35cm 

We are grateful to M. Lublinsky for collaboration at an early stage of the project and valuable discussions throughout the project. We also thank C. Hoyos, D. Kharzeev, K-Y. Kim, F. Pena-Benitez and E. Shuryak for helpful discussions. This work is supported by Alexander von Humboldt Foundation.

\appendix

\vskip 1cm


\begin{thebibliography}{99}

\bibitem{flugra}
  S.~Bhattacharyya, V.~EHubeny, S.~Minwalla and M.~Rangamani,
  JHEP {\bf 0802} (2008) 045
  [arXiv:0712.2456 [hep-th]].\\
  V.~E.~Hubeny, S.~Minwalla and M.~Rangamani,
  arXiv:1107.5780 [hep-th].

\bibitem{haack}
  J.~Erdmenger, M.~Haack, M.~Kaminski and A.~Yarom,
  JHEP {\bf 0901} (2009) 055
  [arXiv:0809.2488 [hep-th]].

\bibitem{indian}
  N.~Banerjee, J.~Bhattacharya, S.~Bhattacharyya, S.~Dutta, R.~Loganayagam and P.~Surowka,
  JHEP {\bf 1101} (2011) 094
  [arXiv:0809.2596 [hep-th]].

\bibitem{surowka}
  D.~T.~Son and P.~Surowka,
  Phys.\ Rev.\ Lett.\  {\bf 103} (2009) 191601
  [arXiv:0906.5044 [hep-th]].

\bibitem{kharzeev}
  D.~Kharzeev,
  Phys.\ Lett.\ B {\bf 633} (2006) 260
  [hep-ph/0406125].\\
  D.~Kharzeev and A.~Zhitnitsky,
  Nucl.\ Phys.\ A {\bf 797} (2007) 67
  [arXiv:0706.1026 [hep-ph]].\\
  D.~E.~Kharzeev, L.~D.~McLerran and H.~J.~Warringa,
  Nucl.\ Phys.\ A {\bf 803} (2008) 227
  [arXiv:0711.0950 [hep-ph]].\\
  K.~Fukushima, D.~E.~Kharzeev and H.~J.~Warringa,
  Phys.\ Rev.\ D {\bf 78} (2008) 074033
  [arXiv:0808.3382 [hep-ph]].

\bibitem{KS}
  D.~E.~Kharzeev and D.~T.~Son,
  Phys.\ Rev.\ Lett.\  {\bf 106} (2011) 062301
  [arXiv:1010.0038 [hep-ph]].

\bibitem{rebhan}
  A.~Rebhan, A.~Schmitt and S.~A.~Stricker,
  JHEP {\bf 1001} (2010) 026
  [arXiv:0909.4782 [hep-th]].\\
  A.~Gynther, K.~Landsteiner, F.~Pena-Benitez and A.~Rebhan,
  JHEP {\bf 1102} (2011) 110
  [arXiv:1005.2587 [hep-th]].

\bibitem{zayakin}
  A.~Gorsky, P.~N.~Kopnin and A.~V.~Zayakin,
  Phys.\ Rev.\ D {\bf 83} (2011) 014023
  [arXiv:1003.2293 [hep-ph]].

\bibitem{kirsch}
  T.~Kalaydzhyan and I.~Kirsch,
  Phys.\ Rev.\ Lett.\  {\bf 106} (2011) 211601
  [arXiv:1102.4334 [hep-th]].

\bibitem{obannon}
  C.~Hoyos, T.~Nishioka and A.~O'Bannon,
  JHEP {\bf 1110} (2011) 084
  [arXiv:1106.4030 [hep-th]].

\bibitem{yee}
  B.~Sahoo and H.~-U.~Yee,
  Phys.\ Lett.\ B {\bf 689} (2010) 206
  [arXiv:0910.5915 [hep-th]].\\
  M.~Torabian and H.~-U.~Yee,
  JHEP {\bf 0908} (2009) 020
  [arXiv:0903.4894 [hep-th]].\\
  D.~E.~Kharzeev and H.~-U.~Yee,
  Phys.\ Rev.\ D {\bf 83} (2011) 085007
  [arXiv:1012.6026 [hep-th]].

\bibitem{neiman}
  C.~Eling, Y.~Neiman and Y.~Oz,
  JHEP {\bf 1012} (2010) 086
  [arXiv:1010.1290 [hep-th]].

\bibitem{pu}
  S.~Pu, J.~-h.~Gao and Q.~Wang,
  Phys.\ Rev.\ D {\bf 83} (2011) 094017
  [arXiv:1008.2418 [nucl-th]].

\bibitem{hou}
  D.~Hou, H.~Liu and H.~-c.~Ren,
  JHEP {\bf 1105} (2011) 046
  [arXiv:1103.2035 [hep-ph]].

\bibitem{logna}
  R.~Loganayagam,
  arXiv:1106.0277 [hep-th].

\bibitem{KY}
  D.~E.~Kharzeev and H.~-U.~Yee,
  Phys.\ Rev.\ D {\bf 84} (2011) 045025
  [arXiv:1105.6360 [hep-th]].

\bibitem{oz}
  Y.~Neiman and Y.~Oz,
  JHEP {\bf 1103}, 023 (2011)
  [arXiv:1011.5107 []].

\bibitem{landsteiner}
  K.~Landsteiner, E.~Megias and F.~Pena-Benitez,
  arXiv:1103.5006 [hep-ph].\\
  I.~Amado, K.~Landsteiner and F.~Pena-Benitez,
  arXiv:1102.4577 [hep-th].\\
  K.~Landsteiner, E.~Megias, L.~Melgar and F.~Pena-Benitez,
  JHEP {\bf 1109} (2011) 121
  [arXiv:1107.0368 [hep-th]].

\bibitem{lin}
  S.~Lin,
  arXiv:1104.5245 [hep-ph].

\bibitem{Min2}
  J.~Bhattacharya, S.~Bhattacharyya, S.~Minwalla and A.~Yarom,
  arXiv:1105.3733 [hep-th].

\bibitem{oz2}
  Y.~Neiman and Y.~Oz,
  JHEP {\bf 1109} (2011) 011
  [arXiv:1106.3576 [hep-th]].

\bibitem{Min}
  J.~Bhattacharya, S.~Bhattacharyya and S.~Minwalla,
  arXiv:1101.3332 [].

\bibitem{Yarom}
  C.~P.~Herzog, N.~Lisker, P.~Surowka and A.~Yarom,
  arXiv:1101.3330 [].

\bibitem{lublin}
  M.~Lublinsky and I.~Zahed,
  Phys.\ Lett.\ B {\bf 684} (2010) 119
  [arXiv:0910.1373 [hep-th]].

\bibitem{nicolis}
  S.~Dubovsky, L.~Hui and A.~Nicolis,
  arXiv:1107.0732 [hep-th].\\
  S.~Dubovsky, L.~Hui, A.~Nicolis and D.~T.~Son,
  arXiv:1107.0731 [hep-th].\\
  A.~Nicolis,
  arXiv:1108.2513 [hep-th].

\bibitem{wz}
  J.~Wess and B.~Zumino,
  Phys.\ Lett.\ B {\bf 37} (1971) 95.

\bibitem{witten}
  E.~Witten,
  Nucl.\ Phys.\ B {\bf 223} (1983) 422.

\bibitem{KRS}
  O.~Kaymakcalan, S.~Rajeev and J.~Schechter,
  Phys.\ Rev.\ D {\bf 30} (1984) 594.

\bibitem{bardeen}
  W.~A.~Bardeen,
  Phys.\ Rev.\  {\bf 184} (1969) 1848.

\bibitem{BPR}
  Y.~Brihaye, N.~K.~Pak and P.~Rossi,
  Phys.\ Lett.\ B {\bf 149} (1984) 191.

\bibitem{isospin}
  D.~T.~Son and M.~A.~Stephanov,
  Phys.\ Rev.\ Lett.\  {\bf 86} (2001) 592
  [hep-ph/0005225].

\bibitem{son}
  D.~T.~Son,
  Phys.\ Rev.\ Lett.\  {\bf 84} (2000) 3771
  [arXiv:hep-ph/9912267].

\bibitem{domain}
  D.~T.~Son and M.~A.~Stephanov,
  Phys.\ Rev.\  D {\bf 77} (2008) 014021
  [arXiv:0710.1084 []].

\bibitem{thompson}
  E.~G.~Thompson and D.~T.~Son,
  Phys.\ Rev.\ D {\bf 78} (2008) 066007
  [arXiv:0806.0367 [hep-th]].

\bibitem{topology}
  D.~T.~Son and A.~R.~Zhitnitsky,
  Phys.\ Rev.\ D {\bf 70} (2004) 074018
  [hep-ph/0405216].

\bibitem{GMS}
  E.~V.~Gorbar, V.~A.~Miransky and I.~A.~Shovkovy,
  Phys.\ Rev.\ C {\bf 80} (2009) 032801
  [arXiv:0904.2164 [hep-ph]].

\end{thebibliography}
\end{document}